\documentstyle[prl,aps]{revtex}
\input epsf
\begin{document}

\title{\bf
Gravitational Lensing Signature of Long Cosmic Strings
}

\author
{Andrew de Laix$^a$, Lawrence M. Krauss$^{a,b}$ and Tanmay Vachaspati$^a$}
\smallskip
\address
{
$^a$Physics Department,
Case Western Reserve University,
Cleveland OH 44106-7079. \\ $^b$Theory Division, CERN, 1211 Geneva, Switzerland
}

\twocolumn[

\maketitle

\begin{abstract}
\widetext

The gravitational lensing by long, wiggly cosmic strings is shown to
produce a large number of lensed images of a background source.  In
addition to pairs of images on either side of the string, a number of
small images outline the string due to small-scale structure on the
string. This image pattern could provide a highly distinctive
signature of cosmic strings.  Since the optical depth for multiple
imaging of distant quasar sources by long strings may be comparable to
that by galaxies, these image patterns should be clearly observable in
the next generation of redshift surveys such as the Sloan Digital Sky
Survey.
\end{abstract}


]

\narrowtext

An exciting outcome of the interplay between particle physics and
cosmology is the realization that topological defects may be present
in our universe and may help to resolve some long-standing puzzles
such as the origin of structure formation. A particular scenario which
has been investigated over the past two decades is one where the
relativistic motion of cosmic strings induces large-scale structure
formation in the wakes that trail behind them \cite{avps}.  Such
strings would be present even now and, if observed, would be a
culminating point for many exciting ideas in particle physics and
cosmology. Past research has mostly focussed on the observation of
cosmic strings by the anisotropies they produce in the microwave
radiation background.  In this paper we consider another promising
means of detecting strings, namely, by observing the lensing of
background galaxies or quasars resulting from strings.

Cosmic strings are lineal gravitating sources with tension along the
string equal to the mass per unit length $\mu$. In the case of strings
produced at the Grand Unified epoch (cosmic time $t \sim 10^{-35}$
secs), $\mu \sim 10^{22}$ gm/cm. The gravitational effects of such
strings are characterized by the dimensionless parameter $8\pi G\mu
\sim 10^{-5}$, implying they are strong enough gravitational sources to
seed structure formation in the universe. A string network consists
of two distinct components: closed loops of string, and long
(infinite) strings. Loops are typically of size $\le \Gamma G\mu t
\sim 10^{-4} t$ and live for about one Hubble time. ($\Gamma \sim 100$
is a numerical factor associated with the rate of gravitational
radiation.)  For infinite strings, numerical simulations of string
networks show that there are order 10 long strings within a horizon
volume at any epoch. These long strings sweep across the horizon at
relativistic speeds and collide and reconnect with other strings. A
very important fact, which has emerged after intensive numerical study
of the string network evolution, is that these long strings are not
straight, but have kinks and wiggles on them \cite{bennett}. The
characterization of this small-scale structure is not fully 
determined \cite{vincent}
especially since the string simulations do not yet include dissipation
of string energy to gravitational radiation. 

We should emphasize that the results given in the literature
\cite{proceeding} describing the string network are ensemble averaged
results.  For gravitational lensing by long strings, however, it will
turn out that the departures from the ensemble averaged string are all
important.  For example, even if the average string has fractal
dimension close to 1, the sharp discontinuities (kinks) on the string
are vital to the gravitational lensing signature of
strings. Unfortunately, no suitable characterization of such
fluctuations is to be found in the current literature, so we were
forced to generate our own ensemble of long strings. The initial
string network is generated by laying down random phases on a periodic
lattice using the method described by Vachaspati and Vilenkin
\cite{tvav}. Then we use the algorithm devised by Smith and Vilenkin
\cite{gsav} to evolve a network of strings in flat spacetime which
allows us to generate a long string sample quickly and efficiently. (A
detailed discussion of the numerical simulation will be presented in a
forthcoming paper\cite{delaix1}.) In using the flat spacetime evolved
strings, our hope is that more realistic string ensembles will also
yield results qualitatively similar to ours.  

Our simulations were run for a time equal to half the light 
crossing time of the box and so the periodic boundary effects did 
not become significant. Segments
of string within a cube with sides half the length of
the box have had sufficient time to relax to a constant, fractal
structure, and it is these segments which we use for lensing
calculations since longer segments retain the random walk
structure of the initial conditions.

After generating a long string, the next step is to find photon
trajectories in its gravitational field. The location of the string at
time $t$ is given by a vector function ${\bf f} (\sigma ,t)$ where
$\sigma$ is a parameter along the string. In the absence of the
strings, photons would travel from the source to the observer along
the trajectory described by their constant initial four velocity
$\gamma^\mu$.  When a string is present, the photons are deflected,
but, as $8\pi G\mu$ is very small, we need only calculate the
bending to first order in this parameter, that is, in the weak
field approximation where we write the metric as $g_{\mu \nu} =
\eta_{\mu \nu} + h_{\mu \nu}$.  In our work on lensing by string loops 
\cite{delaix}, we had shown that for any compact lens, even those moving 
at relativistic speeds, the photon deflection can be reduced to a
single integral along the string on the time slice:
\begin{equation}
t_{0} = f_{\|}(\sigma, t_{0})
\label{timeslice}
\end{equation} 
where $f_\|$ is the component of $\bf f$ that is parallel to the
photon trajectory.  To derive the deflection formula in \cite{delaix},
it was assumed that the source and observer reside at distances much
larger than the size of the loop, but in the case of long strings, the
source and observer are separated by about the horizon size, while the
length of string which contributes to the gravitational potential is
equally long.  We can show, however, that for photons passing near
the string, the dominant effect will be produced by only a small
fraction of the overall string \cite{delaix1}. 
So let us consider the deflection
arising from a finite segment of an infinite string over which
$\sigma$ ranges from $\sigma_1$ to $\sigma_2$. In this case the
deflection angle is
\begin{equation}
\bar{\alpha_i} = -4 G\mu 
\int_{\sigma_1}^{\sigma_2} d \sigma \left [
\frac{F_{\mu \nu}(\sigma, t)
\gamma^{\mu} \gamma^{\nu}}
{1 - \dot{f}_{\|}}\frac {{f}_{\bot i}} {f_{\bot}^2}
\right ]_{t = t_0} 
                \label{deflection}
\end{equation}
where, ${\bf f} (\sigma ,t) = {\bf f}_\bot + {\bf f}_\|$, and, ${\bf
f}_\bot$ is the component of $\bf f$ which is perpendicular to the
photon trajectory. The index $i$ on $\alpha$ takes on two values to
denote components in the plane perpendicular to $\vec \gamma$.  The
tensor $F_{\mu \nu}$ is given by
\begin{equation}
        F_{\mu \nu} = \dot{f}_{\mu}\dot{f}_{\nu} - f_{\mu}' f_{\nu}' -
        \eta_{\mu \nu}\dot{f}^{2}.
        \label{stress}
\end{equation}
where overdots and primes refer to derivatives with respect to $t$ and
$\sigma$. 
%
%
%
%

In eq. (\ref{deflection}) we have also discarded a boundary term
whose effect on the image distortion is suppressed by the ratio of the
size of a galaxy to the distance between the galaxy and the string.
We assume that the galaxy is at a large distance from the 
string and are justified in ignoring this contribution to
${\bar \alpha}$.

%
%
To calculate the photon deflection in eq. (\ref{deflection}), we need
to find the string coordinates on the time slice given in
eq.\ (\ref{timeslice}). For this, we took the unperturbed photon
trajectory to be parallel to the z axis and found $f_\|$ at all times
using our long string evolution code.  This allows us to numerically solve
eq.\ (\ref{timeslice}) and obtain the string trajectory at the
particular time slice needed in eq.\ (\ref{deflection}). 
All our string sections had fractal
dimension of about 1.25 within the scale of our 256 by 256 box and
above the numerical cut-off scale $\Delta \sigma$ which we set 
to correspond to a physical scale given by $\Delta \sigma/D_l = 0.1''$ 
where $D_l$ is the angular diameter distance measured from the observer 
to the string segment. This choice of $\Delta \sigma$ is small enough
for discretization effects to be unimportant in our study of lensing
and large enough that the string segment we have constructed stretches
across a big enough patch of the sky. (The full 256 by 256 box
corresponds to a 25'' by 25'' square patch of the sky.)
%
%

To see how the deflection in eq.\ (\ref{deflection}) causes lensing,
consider an axis connecting the observer to the lens, and suppose that
there is an image located at a small angular displacement ${\bf x}$
from the axis.  If we define
\begin{equation}
        {\bf \alpha} = \bar{{\bf \alpha}} {D_{ls} \over D_{s}},
        \label{alpha2}
\end{equation}
where $D_{ls}$ is the angular diameter distance from the lens to the
source and $D_{s}$ is the angular diameter distance from the observer
to the source, then the angular location of the source, ${\bf y}$,
which would produce this image is given by
\begin{equation}
        {\bf y} = {\bf x} - {\bf \alpha}({\bf x}),
        \label{y}
\end{equation}
in the limit of small angles which is always valid here.  We have also
assumed the thin lens approximation in which the dimensions of the
lens are much smaller than $D_{ls}$ or $D_{s}$.  This is valid as long
as we consider short segments of string when compared to the horizon
scale.  

Before describing our results, it is worthwhile to point out the
length scales in the problem. We consider two possible sources for
lensing: quasars and high redshift galaxies.  Quasars are very
luminous, point--like objects with typical redshifts of about $z \sim
2-5$.  The typical angular separation of quasars is much larger than
that of the typical image separation, so lensed pairs will be
distinct. Another set of objects worth observing are galaxies at a
redshift of $z \sim 2-3$, with the visible portion of such galaxies
having angular diameters of roughly 0.5''.  The angular separation for
the galaxies is about 10''\cite{sawicki}.  The angular diameter of the
length scale associated with small-scale structure on the string is
$\Gamma G\mu t$, which, for a string at a redshift of $z
\sim 1$, 
is very similar to the angular separation of high redshift galaxies
corresponding to an angular scale $\sim 10$'' for $\Gamma = 100$.
Finally, note also
that the mass associated with small scale string structure is 
$O(10 M_{gal})$, so that multiple lensed images could have 
separations on the order of 10'', compared to multiple imaging by 
galaxies, which generally yields image separations of 1-3''.

\begin{figure}[tbp]
\caption{ \label{quasar} Several quasar lensing systems.  The hatched circle
shows the location of the unlensed source and the open circles show the
location of the resulting images produced by the string segment
(dotted line).  The ratio of the areas gives the relative
magnification to the source.}
\epsfxsize = \hsize \epsfbox{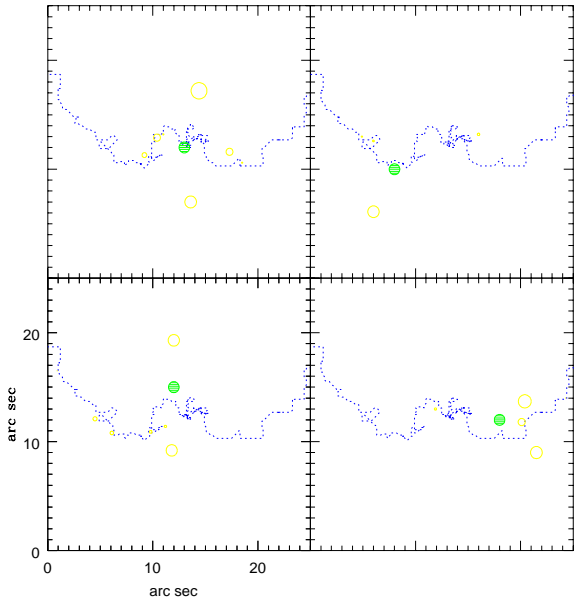}
\end{figure}

First we shall consider the lensing of a quasar due to a long cosmic
string. In Fig. \ref{quasar}, we show string lensing for four
different quasars each located at a redshift of $z = 2$.  The
projected string configuration is located at $z = 1$, and is shown by
the dotted line, where only the contribution from the string shown was
used in calculating the image locations.  The hatched circle shows the
location of the unlensed source, while the open circles show the
locations of the various images, where the areas of the open circles
are proportional to the magnification of the images relative to the
source.

\begin{figure}[tbp]
\caption{ \label{crit2} The projected string configuration along with
the unlensed sources.}
\epsfxsize = \hsize \epsfbox{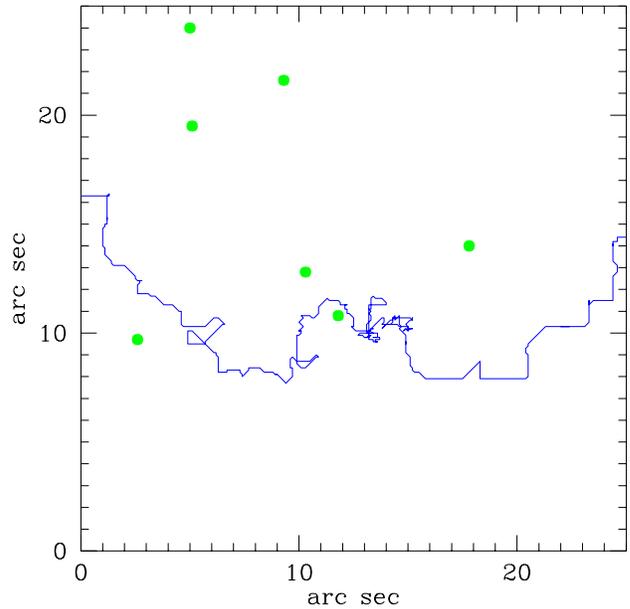}
\end{figure}

\begin{figure}[tbp]
\caption{ \label{image2} The resulting images of the sources shown in
Figure \protect\ref{crit2}.  Only the string points shown in Figure
\protect\ref{crit2} were used in determining the lensing effects.}
\epsfxsize = \hsize \epsfbox{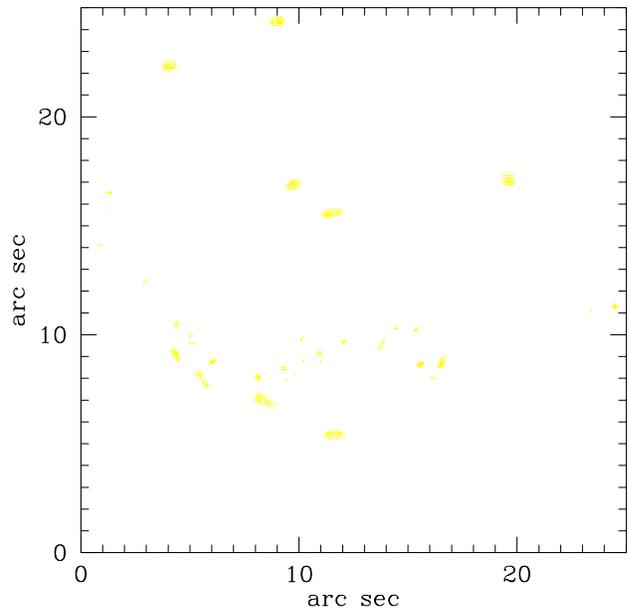}
\end{figure}

In Fig. \ref{crit2} we show the same projected segment of string as
used for the quasar lensing along with the unlensed images of seven
circular sources located at redshift $z = 2$.  The locations of these
sources are chosen randomly in the plane and they have roughly the
proper size and density to correspond to high redshift galaxies.  In
Figure \ref{image2} we see the images of these sources as they would
appear to a terrestrial observer. (We have set $G\mu = 10^{-6}$ in the
simulations.)  Note that the 0.1'' resolution is consistent with that
which can be obtained from the Hubble space telescope, while ground
based telescopes achieve resolutions of about 0.5''.  

It is clear from Figs. \ref{quasar} \& \ref{image2} that a sequence
of small demagnified images outline the string. One can qualitatively
understand this result by considering a chain of point masses
separated by distances less than their Einstein radius.  Between the
masses, the gravitational deflection can be cancelled by the opposite
attraction of the two masses.  This allows the formation of a
small image near the axis connecting the two masses, and for a chain, one
might expect to see several small images. In the case of a string, it is the
wiggles and kinks which provide the breaks needed to form the small
images.  The effect is similar to that observed for open string loops
\cite{delaix}; photons passing through a kink are subject to  a
diminished deflection which can allow an image to be formed.  
\newpage


Thus, the lensing pattern by strings has two distinct signatures: the
first is a sequence of baby images that outline the string and the
second is a sequence of major images that lie on either side of the
string. The first feature depends on the frequency and magnitude of
small-scale structure of the string and can be highly distinctive of
the stringy nature of the lens.  The second feature does not depend on
the small-scale structure of the string and is similar to the
previously considered lensing by straight strings due to the conical
nature of space-time that they produce.
  
It should be noted, however, that while the galaxy images in Figure
\ref{image2} are striking, it is not so clear that actual observed
galaxy lensing by strings would be quite so distinctive. In the first
place, the galaxy brightnesses are much smaller, so that the small
images near the string may not be visible.  Secondly, due to the
higher space density of galaxies, image separations due to string
lensing will be comparable to the separations of galaxies themselves.
For both of these reasons, we suggest that a survey involving quasar
lensing is probably the most efficient first step in any attempt to
probe for lensing due to cosmic strings.  As we describe below,
galaxies can then provide a useful second probe to confirm the string
origin of any quasar lensing.

With that in mind, we consider the probability of observing lensing by
long cosmic strings and strategies for observation. As described
earlier, it is most useful to first consider surveys of objects which
are widely separated on arc second scales, such as quasars. Given that
the string density is on the order of ten per horizon volume, the
probability of a given high redshift object being lensed by a long
string has been estimated to be about $\tau \approx 100 G\mu {\rm ln}
(1+z) \sim 10^{-4}$\cite{avps}. A more refined calculation, based in
part on estimates of the long string density arising from numerical
simulations\cite{proceeding}, and also on the projected angular cross
section for string lensing yields an optical depth of $\tau \approx
1. \times 10^{-3}$ ( see \cite{delaix1} for further details).  To
turn this into an actual prediction of lensing frequency in any
specific survey requires an analysis of such issues as selection
effects, etc.\cite{TOG}.  However, it is useful to compare this
optical depth to that due to lensing of high redshift quasars by the
known galaxy population.  In the ratio of these quantities many such
effects should cancel. For a flat universe, the optical depth,
assuming the dominant lensing is by elliptical galaxies modeled as
isothermal spheres, is $\approx 3\times 10^{-3}$
\cite{TOG,krausswhite}.  If account is taken of possible finite core
effects\cite{hinshkrauss}, this number could be reduced by a factor of
perhaps two.  Hence, the ratio of optical depths for string lensing
and galaxy lensing in any large sample of quasars is of order unity,
suggesting on average as many string induced multiply imaged quasars
as galaxy induced ones.

Existing surveys have unearthed on the order of a dozen multiply
imaged quasars, so one might expect that there should be several
string candidates in this sample. However, because the long string
density in our horizon volume is small, a significant fraction of the
sky would have to be surveyed before a definitive constraint could be
derived (i.e. the string sample is non-Gaussian). A good search
strategy would be to observe a large number of quasars in a wide
angular field, and the Sloan Digital Sky Survey is precisely such an
observation.  Approximately $10^5$ quasars will be observed over 1/4
of the sky with typical redshifts of about $z \sim 2$.  One can expect
that the SDSS will observe at least on the order of several hundred
lensed quasars, with roughly equal numbers due to galaxies
and long strings.  If the strings have significant small scale
structure, or induce anomalously large angular splittings, one may
expect to distinguish the two types of events when the lensing galaxy
is not visible, but one cannot necessarily be certain.  Here, though,
is where galaxy observations can be useful.  By using a sensitive
telescope like the Hubble Space Telescope, one can observe the high
redshift galaxies in the neighborhood of the lensed quasar and look
for evidence of string lensing of galaxies.  Together with the quasar
images, these would likely provide incontrovertible evidence of cosmic
strings, or by their absence, rule them out as seeds of structure
formation.

We would like to thank Glenn Starkman and Heather Morrisson for 
discussions, and the Department 
of Energy for support.



\begin{thebibliography}{999}

\bibitem{avps} A. Vilenkin and E.P.S. Shellard, ``Cosmic Strings and
Other Topological Defects'', Cambridge University Press (1994).

\bibitem{bennett} 
D.P. Bennett and F.R. Bouchet, Phys. Rev. {\bf D41}, 2048 (1990);
B. Allen and E.P.S. Shellard, Phys. Rev. Lett. {\bf 64}, 119 (1990). 

\bibitem{vincent} For a recent discussion, see G.R. Vincent, M. Hindmarsh
and M. Sakellariadou, SUSX-TH-96-020, astro-ph/9612135.

\bibitem{proceeding} See the papers by D.P. Bennet, F.R. Bouchet,
N.Turok, A. Albrecht, and, E.P.S. Shellard and B. Allen in 
{\it The Formation and Evolution of Cosmic Strings}, eds. 
G.W. Gibbons, S.W. Hawking and T. Vachaspati,
(Cambridge University Press 1990).

\bibitem{tvav} T. Vachaspati and A. Vilenkin, Phys. Rev. {\bf D30}, 
2036 (1984).

\bibitem{gsav} A. G. Smith and A. Vilenkin, Phys. Rev. {\bf D36}, 990
(1987).

\bibitem{delaix1} A. A. de Laix, in preparation.

\bibitem{delaix} A. A. de Laix and T. Vachaspati, Phys. Rev. D {\bf
54}, 4780 (1996).

\bibitem{sawicki} M. J. Sawicki, H. Lin and H. K. C. Yee, preprint
astro-ph/9610100.

\bibitem{TOG} E. L. Turner, J. P. Ostriker and J. R. Gott, Astrophys. J.
{\bf 284}, 1 (1984)

\bibitem{krausswhite} L. M. Krauss and M. White, Astrophys. J. {\bf
394}, 385 (1992)

\bibitem{hinshkrauss} G. Hinshaw and L. M. Krauss,
Astrophys. J. {\bf 320}, 468 (1987)


\end{thebibliography}
\end{document}